\documentclass[superscriptaddress,twocolumn,10pt,prl,floatfix,nofootinbib,aps]{revtex4-2} 

\newif\ifpreprint
\preprinttrue

\usepackage[labeled]{multibib}

\newcites{Methods}{Methods section}

\input{glyphtounicode}
\usepackage[T1]{fontenc}
\usepackage[utf8]{inputenc}
\usepackage[english]{babel}
\usepackage{amsmath}  
\usepackage{amsfonts} 
\usepackage{amssymb}
\usepackage{graphicx} 
\usepackage{float}  
\usepackage{xcolor}
\usepackage{placeins}
\usepackage{hyperref}
\usepackage[scaled]{helvet}

\hypersetup{
    bookmarksopen,bookmarksnumbered,
    colorlinks,
    linkcolor=blue,
    citecolor=blue
    }
\graphicspath{ {./} }
\setcitestyle{super}
\usepackage{array}
\usepackage[table]{xcolor}
\usepackage{braket}
\usepackage{booktabs}
\usepackage{multirow}
\usepackage{dsfont} 
\usepackage[justification=justified,singlelinecheck=false]{caption}

\usepackage{tabularx}
\usepackage{siunitx}

\usepackage{subcaption}
\usepackage{bigstrut}
\usepackage{svg}
\usepackage{upgreek}

\usepackage{newfloat}

\usepackage{lineno}

\DeclareFloatingEnvironment[name={\textbf{Supplementary Figure}}]{suppfigure}

\newcommand{\aref}[1]{\hyperref[#1]{Appendix~\ref*{#1}}}
\DeclareCaptionLabelSeparator{line}{\hspace{2pt}\bf{|}\hspace{2pt}}

\DeclareFloatingEnvironment[name={\textbf{Extended Table}}]{supptable}

\definecolor{reviewercolor}{RGB}{0,0,0}

\newcommand{\change}[1]{%
\textcolor{reviewercolor}{\ignorespaces#1\unskip}%
}

\begin{document}

\captionsetup[table]{
    name ={\bf{Table}},
    labelsep=period,
    justification=raggedright,
    font=small,
    singlelinecheck=false
    }
\captionsetup[figure]{
    name={\bf{Figure}},
    labelsep=line,
    justification=raggedright,
    font=small,
    singlelinecheck=false
    }

\captionsetup[supptable]{
    name={\bf Extended Table},
    labelsep=line,
    justification=raggedright,
    font=small,
    singlelinecheck=false
}

\renewcommand{\equationautorefname}{Eq.}
\renewcommand{\figureautorefname}{Fig.}
\renewcommand*{\sectionautorefname}{Sec.}

\title{Eight-Qubit Operation of a 300 mm SiMOS Foundry-Fabricated Device}

\author{\textbf{Andreas Nickl}$^{+,*}$}
\affiliation{School of Electrical Engineering and Telecommunications, University of New South Wales, Sydney, NSW 2052, Australia}

\author{Nard Dumoulin Stuyck}
\affiliation{School of Electrical Engineering and Telecommunications, University of New South Wales, Sydney, NSW 2052, Australia}
\affiliation{Diraq, Sydney, NSW, Australia}
\author{Paul Steinacker}
\affiliation{School of Electrical Engineering and Telecommunications, University of New South Wales, Sydney, NSW 2052, Australia}
\affiliation{Diraq, Sydney, NSW, Australia}
\author{Jesus D. Cifuentes}
\affiliation{School of Electrical Engineering and Telecommunications, University of New South Wales, Sydney, NSW 2052, Australia}
\affiliation{Diraq, Sydney, NSW, Australia}
\author{Santiago Serrano}
\affiliation{School of Electrical Engineering and Telecommunications, University of New South Wales, Sydney, NSW 2052, Australia}
\affiliation{Diraq, Sydney, NSW, Australia}

\author{MengKe Feng}
\affiliation{School of Electrical Engineering and Telecommunications, University of New South Wales, Sydney, NSW 2052, Australia}
\affiliation{Diraq, Sydney, NSW, Australia}

\author{Ensar Vahapoglu}
\author{Fay E. Hudson}
\author{Kok Wai Chan}
\affiliation{School of Electrical Engineering and Telecommunications, University of New South Wales, Sydney, NSW 2052, Australia}
\affiliation{Diraq, Sydney, NSW, Australia}
\author{Stefan Kubicek}
\author{Julien Jussot}
\author{Yann Canvel}
\author{Sofie Beyne}
\author{Yosuke Shimura}
\affiliation{Imec, Leuven, Belgium}
\author{Roger Loo}
\affiliation{Imec, Leuven, Belgium}
\affiliation{Department of Solid-State Sciences, Ghent University, Krijgslaan 285, 9000 Ghent, Belgium}
\author{Clement Godfrin}
\author{Bart Raes}
\author{Sylvain Baudot}
\author{Danny Wan}
\affiliation{Imec, Leuven, Belgium}
\author{Arne Laucht} 
\affiliation{School of Electrical Engineering and Telecommunications, University of New South Wales, Sydney, NSW 2052, Australia}
\affiliation{Diraq, Sydney, NSW, Australia}
\author{Chih-Hwan Yang}
\affiliation{School of Electrical Engineering and Telecommunications, University of New South Wales, Sydney, NSW 2052, Australia}
\affiliation{Diraq, Sydney, NSW, Australia}
\author{Wee Han Lim}
\affiliation{School of Electrical Engineering and Telecommunications, University of New South Wales, Sydney, NSW 2052, Australia}
\affiliation{Diraq, Sydney, NSW, Australia}
\author{Andre Saraiva}
\affiliation{Diraq, Sydney, NSW, Australia}
\author{Christopher C. Escott}
\affiliation{Diraq, Sydney, NSW, Australia}
\author{Kristiaan De Greve}
\affiliation{Imec, Leuven, Belgium}
\affiliation{Proximus Chair in Quantum Science and Technology and Department of Electrical Engineering (ESAT-MNS), KU Leuven, Leuven, Belgium}
\author{Andrew S. Dzurak}
\author{Tuomo Tanttu$^{*,}$}
\affiliation{School of Electrical Engineering and Telecommunications, University of New South Wales, Sydney, NSW 2052, Australia}
\affiliation{Diraq, Sydney, NSW, Australia}

\date{\today}

\begin{abstract}
\boldmath

\unboldmath
Silicon spin qubits are a promising platform for quantum computing due to their high coherence, controllability, and CMOS manufacturability, yet scalable implementations have so far been limited to a few qubits.
Here, to take a step towards large\change{r qubit} systems, we tune and coherently control an eight-dot linear array of silicon spin qubits fabricated in a  \SI{300}{\milli \meter} CMOS-compatible foundry process, establishing operational scalability beyond the two-qubit regime.
All eight qubits are successfully tuned and characterized as four double-dot pairs, exhibiting Ramsey dephasing times $T_2^*$ up to \SI{41(2)}{\micro \second} and Hahn-echo coherence times $T_2^{\mathrm{Hahn}}$ up to \SI{1.31(4)}{\milli \second}.
Readout of the central four qubits is achieved via a cascaded charge-sensing protocol, enabling high-fidelity measurements of the entire multi-qubit array in a two step process.
Additionally, we demonstrate a two-qubit gate operation between adjacent qubits with low phase noise. We show that silicon spin qubit arrays can be scaled to medium-sized arrays of 8 qubits while maintaining system coherence.

\end{abstract}
\maketitle

\def\thefootnote{*}
\footnotetext{Correspondence to: \texttt{a.nickl@unsw.edu.au}, \texttt{t.tanttu@unsw.edu.au}}
\def\thefootnote{\arabic{footnote}}

\section{Introduction}
\noindent
Scaling quantum processors from few-qubit demonstrations to viable devices for fault-tolerant quantum computing depends critically on the ability to fabricate, tune, and coherently control larger arrays of qubits with industrially relevant metrics~\cite{taylor_fault-tolerant_2005, kane_silicon-based_1998, kim_evidence_2023, acharya_quantum_2025}.  
Among the leading platforms, silicon spin qubits in quantum dots stand out due to their compatibility with CMOS manufacturing~\cite{veldhorst_silicon_2017, gonzalez-zalba_scaling_2021, zwerver_qubits_2022}, long coherence times in isotopically purified materials, and the possibility of integrating control and readout circuitry in a scalable architecture~\cite{bartee_spin-qubit_2025, stuyck_cmos_2026, thomas_rapid_2025}. Linear spin qubit arrays have been demonstrated in Ge\cite{hendrickx_four-qubit_2021, jirovec_mitigation_2025, john_robust_2025} and Si/Ge\cite{philips_universal_2022, fernandez_de_fuentes_running_2026, ha_two-dimensional_2025, undseth_weight-four_2026, tidjani_three-dimensional_2025} quantum dots, while CMOS-based implementations have so far been limited to one or two qubits\cite{stuyck_demonstration_2024, huang_high-fidelity_2024, steinacker_bell_2025, steinacker_industry-compatible_2025, yang_operation_2020}.
\change{
    While fault-tolerant quantum computing will likely require extending beyond strictly linear arrays to multi-row or two-dimensional architectures \cite{ha_two-dimensional_2025, siegel_towards_2024, boter_spiderweb_2022}, linear arrays provide a simple and well-controlled platform for developing and benchmarking qubit control techniques.
    }

Quantum dot spin qubits fabricated in a \SI{300}{\milli \meter} CMOS process can yield low and consistent charge noise across devices~\cite{elsayed_low_2024} with single- and two-qubit gate fidelities exceeding \SI{99}{\percent}~\cite{stuyck_demonstration_2024, steinacker_industry-compatible_2025}. These results demonstrate that the materials, gate stack engineering, and fabrication uniformity required for scaling are already viable. Yet, they remain limited to low qubit counts and do not fully address device variability or coherent operation of qubits across greater linear arrays.

In this work, we build on the same \SI{300}{\milli \meter} CMOS-compatible fabrication process developed for prior experiments~\cite{stuyck_demonstration_2024, steinacker_industry-compatible_2025} and extend the device and control methodology to an eight electron spin qubit linear array. We tune all eight dots, characterize single-qubit coherence ($T_2^*$ and $T_2^{\text{Hahn}}$) across the array, and demonstrate the feasibility of two-qubit gate operations among a pair of qubits. We implement a cascaded charge-sensing architecture for the central four qubits to permit simultaneous high-fidelity readout within the extended linear chain \cite{van_diepen_electron_2021, chittock-wood_radio-frequency_2025}. These results demonstrate operational scalability in silicon spin qubits beyond the two-qubits, showing that the fabrication, control, and readout techniques developed in small devices can be translated to larger linear arrays in CMOS-compatible platforms.

\section{Results}
\subsection{Device}
\noindent
\ifpreprint
\begin{figure*}
    \includegraphics[width=\textwidth, angle = 0]{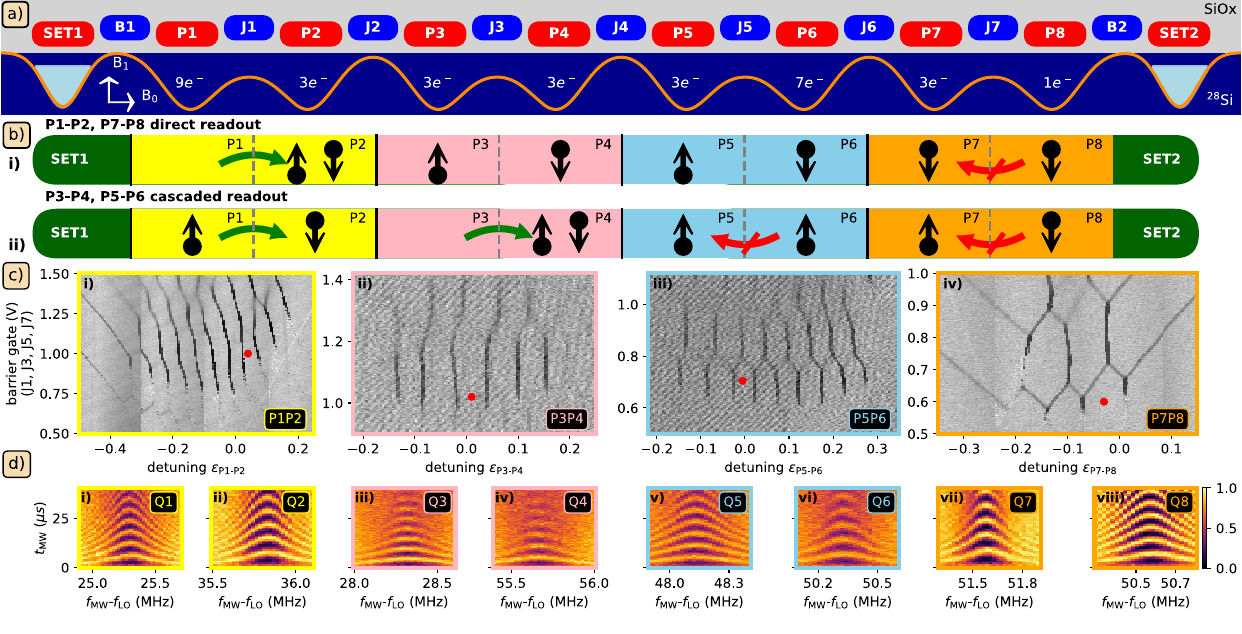} 
    \caption{
        \textbf{Overview of operation and calibration of 8 dot device.}
        \textbf{a)} Schematic of cross section of the device depicting silicon ($^{28}$Si) substrate, oxide layers (shades of gray), and plunger electrodes (P$_i$ and SET$_i$) and barrier (J$_i$ and B$_i$) gates. The electric potential is visualized within the Si substrate with the used electron occupancy.
        \change{A SEM scan of a similar device is shown in Supplementary Figure 7.} 
        \textbf{b)} Spin-to-charge conversion readout techniques of the lateral DQDs: i) P1-P2 and P7-P8 via direct SET readout and the central DQDs; ii) P3-P4 and P5-P6 via cascaded readout facilitated by electrons in lateral dots. The left side exemplifies charge movement for odd spin states (green arrows), while the right side exemplifies Pauli spin blockade for even spin states (red arrows).
        \textbf{c)} Charge stability maps in isolated mode for i) P1--P2, ii) P3--P4, iii) P5--P6, and iv) P7--P8. Red circles mark the charge configuration used for the measured qubits
        \textbf{d)} Rabi-chevron measurements for qubits 1-8, shown in i)-viii) respectively.
    }
    \label{fig:mega_figure}
\end{figure*}
\fi
The device consists of a linear array of eight quantum dots, with single-electron transistors (SET) integrated at both ends for spin readout (Fig.~\ref{fig:mega_figure}a). The design and fabrication were performed by imec with a \SI{300}{\milli \meter} silicon MOS fabrication workflow on a $^{28}$Si wafer, optimized for low noise and defect densities with a residual amount of \SI{400}{ppm} of $^{29}$Si \cite{elsayed_low_2024,li_flexible_2021}.
\change{
    The gate geometry and pitch (\SI{90}{\nano\meter}) used here were chosenyeah  as an evolutionary step in the development of quantum dot fabrication techniques.
    These parameters can be further optimized for the formation of MOS electron spin qubits by bringing the quantum dots closer together \cite{steinacker_industry-compatible_2025, jones_mid-circuit_2026}.
    This improves electrostatic confinement and enables more robust tuning of tunnel coupling and exchange interactions between neighboring qubits \cite{cifuentes_bounds_2024}.
}

The device is operated as four unit cells of two qubits per double quantum dot (DQD) which are captured under two neighboring plunger gates (P) each. This technique breaks down the complex task of forming an 8 qubit system to forming mostly independent well-understood two-qubit systems. Oddly numbered barrier gates (J) control the intra-DQD tunneling while evenly indexed barrier gates determine the inter-DQD tunnel coupling. A schematic of the gate geometry can be seen in Fig.~\ref{fig:mega_figure}a.

Gate electrodes are made from poly-crystalline silicon to minimize lattice strain compared to historically used aluminum gates \cite{huang_high-fidelity_2024,veldhorst_two-qubit_2015} and are electrically insulated from each other by an oxide layer.
The electrons are simultaneously loaded from the two-dimensional electron gas (2DEG) formed by the lateral SETs, either directly to the adjacent DQDs (P1--P2 or P7--P8) or through them for the two central DQDs P3--P4 and P5--P6. We form a continuous 2DEG from the SET island to the desired dot pair and subsequently raised the dot potential until the desired electron number is acquired. The charge occupations shown in Fig.~\ref{fig:mega_figure}c are chosen to achieve the best possible qubit operating regime. A complete schematic of the electron loading routine, providing independent access to all charge occupations (see Supplementary Figure 2), is shown in the Supplementary Figure 1.
\change{
    We expect this technique to be applicable to longer 1D arrays but suggest loading only the edge dots with electrons and shuttling to inner dots for more complex devices to avoid a full reset of charge occupancies.
}

\change{
    The electrons in each dot form an effective spin-half system that is individually controlled via electron spin resonance (ESR) using a TiN stripline microwave-antenna located above the mentioned gate electrodes applying an oscillating magnetic field $B_1$. 
    This field is engineered to be homogeneous across the entire array and oscillates out-of-plane as shown in the schematic in Fig.~\ref{fig:mega_figure}a.
}
Dot pairs from P1-P2 to P7-P8 have respective electron configurations of (9-3), (3-3), (3-7) and (3-1).
\change{
    Larger quantum dots exhibit lower excited state energies, increasing the influence of surface roughness at the silicon-oxide interface towards dot formation.
    The electron wavefunction can be modified either by adjusting the electrostatic confinement potential or by changing the electron occupation number.
    This sensitivity to tuning parameters leads to the observed variation in usable charge occupancies, which may appear arbitrary.
    Additional charge configurations explored in this work are summarized in Supplementary Table 1.
}
All measurements are performed in a $^3$He/$^4$He dilution refrigerator operated at a base temperature of $\sim$\SI{20}{\milli \kelvin} with a vector magnet.

\subsection{Operation}
\noindent
We initialize pairwise spin parity states by detuning the plunger gate voltages to an even-even charge distribution (e.g: from 9-3 to 10-2 in P1--P2) and waiting for \SI{100}{\micro \second} to allow the spin pair to decay to its ground state $\ket{\uparrow\downarrow}/ \ket{\downarrow\uparrow}$. By ramping back to the operational (e.g., 9-3 in P1--P2) odd-odd charge configuration diabatically, one can initialize the mixture of odd states $\ket{\downarrow\uparrow}$ and $ \ket{\uparrow\downarrow}$, respectively.
Pure spin states are initialized by ramping to a $T_1$ decay-hotspot within the qubits odd--odd electron configuration and confirming success by collapsing the wavefunction to the $\ket{\downarrow\downarrow}$ state by measurement via a heralding protocol~\cite{yang_operation_2020, huang_high-fidelity_2024}.
We utilize Pauli spin blockade (PSB) to read the qubit pairs' spin parity state~\cite{johnson_singlet-triplet_2005, seedhouse_pauli_2021}.
Polarized triplet states prevent charge movement from dot 1 to dot 2 when ramping to the PSB region, whereas unpolarized parity states are free to tunnel. Those charge movements are captured by the SET~\cite{seedhouse_pauli_2021}. All electron occupations during the described operating regimes are given in Tab.~\ref{tab:e_occupations}.

\begin{table}
\centering
    \begin{tabular}{|l|c|c|c|c|}
    \hline
    \textbf{$\mathbf{e^-}$ occupation:} & \textbf{P1-P2} & \textbf{P3-P4} & \textbf{P5-P6} & \textbf{P7-P8} \\
    \hline
    
    qubit control        & (9-3) & (3-3) & (3-7) & (3-1) \\
    initialization \& readout & (10-2) & (4-2) & (4-6) & (4-0) \\
    \hline
    \end{tabular}
\caption{
    Double quantum dot electron occupations during control, initialization and readout. \label{tab:e_occupations}
}
\end{table}
The central four dots P3--P6 are measured via electron cascading where PSB readout is performed, but the lateral sub-systems are tuned to be close to their electron anti-crossing while being in an unblockaded spin state. A charge movement in P3--P4 or P5--P6 triggers an electron cascade in P1--P2 or P7--P8, respectively,  which is then read out with an increased SNR by the SETs compared to directly sensing the central part of the device \cite{van_diepen_electron_2021}.  This cascaded readout scheme is designed to keep the number of electrons in each DQD constant. Figure~\ref{fig:mega_figure}b-ii provides a schematic of unblockaded PSB readout cascade on the left half of the device and the blockaded on the right half. Calibration and the difference in visibility is shown in Supplementary Figure 4.

Spin up and down states are split by the Zeeman energy $\sim$\SI{14}{\giga \hertz} ($\sim$\SI{58}{\micro \electronvolt}) due to an external in-plane DC magnetic field of $B_0=$\SI{0.5}{\tesla}. Small differences in electron g-factors allow direct addressability of all qubits with separate resonant ESR pulses \cite{veldhorst_spin-orbit_2015}.
Single qubit gates $X_{\pi/2}$ are realized by resonantly exciting the electrons with a timed microwave pulse, while $Z_{\pi/2}$ gates applied by a virtual phase shift in the microwave source~\cite{vandersypen_nmr_2005}. The Heisenberg exchange interaction between two neighboring qubits is controlled via base band control of the barrier gates J by which controlled phase gates (CZ) are realized \cite{veldhorst_two-qubit_2015, tanttu_assessment_2024, watson_programmable_2018}.
The SET top-gate operation voltage and the qubits Larmor frequencies are being tracked and corrected by real-time feedback protocols \cite{dumoulin_stuyck_silicon_2024}.

All measurements were performed with the given electron numbers in each dot provided in Tab.~\ref{tab:e_occupations} with results of the lateral double dots being acquired simultaneously. The same is true for the characterization of the two central pairs.
\subsection{Qubit Characterization}
\ifpreprint
\begin{figure}
    \includegraphics[width=1.0\linewidth, angle = 0]{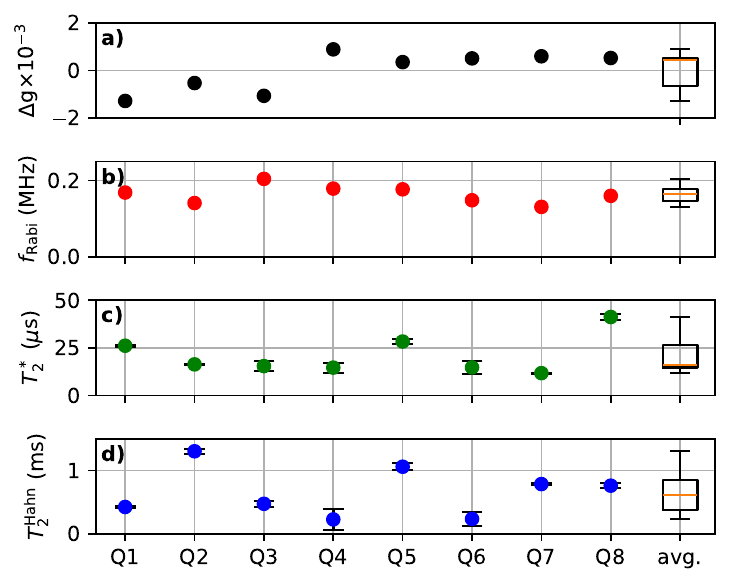}
    \caption{\textbf{Qubit characteristics summary.}
        \textbf{a)} Larmor frequencies, 
        \textbf{b)} Rabi frequencies,
        \textbf{c)} Ramsey coherence times $T_2^*$,
        \textbf{d)} Hahn echo coherence times $T_2^{\rm Hahn}$.
        The box symbol shows the spread of values as well as their mean and standard deviation.
    }
    \label{fig:summary}
\end{figure}
\fi

\noindent 
All eight quantum dots were tuned to accommodate a qubit each by adjusting gate voltages and operational parameters for initialization, control and readouts. Rabi chevrons of all measured qubits are shown in Fig.~\ref{fig:mega_figure}d.
The extended top-gate dimensions and the pitch to their neighbors did not facilitate the most obvious electrostatic gate voltage configuration by applying a similar voltage on all gates. Significantly more negative potentials were applied to evenly indexed barrier gates in order to spatially confine electron pairs enabling the formation of qubits.

The qubits' g-factors are distributed over $\Delta g = 2.17\times10^{-3}$ of each other, allowing individual addressability via ESR drive. Figure~\ref{fig:summary}a shows their individual, relative g-factors. 
A larger spread of $\Delta g = 7\times10^{-3}$ has been predicted from atomistic tight binding simulations in a similar platform\cite{cifuentes_bounds_2024}. The difference is likely to be due the larger dot sizes in this work. Similar Larmor frequencies are particularly interesting for global control techniques where qubits are constantly driven to allow scalable qubit base-band control and decouple qubits from quasi-static noise \cite{seedhouse_quantum_2021, hansen_entangling_2024}.
Further, variations of g-factors can be modified by changing the angle of $B_0$ as shown in Supplementary Figure 6~\cite{tanttu_controlling_2019, jock_silicon_2018, cifuentes_bounds_2024}.

Qubit Rabi frequencies are all in range 141(1)--\SI{204.5(6)}{\kilo \hertz} (Fig.~\ref{fig:summary}b). This together with the lack of significant Rabi frequency change as a function of gate voltages hints towards mostly magnetic drive through the used line antenna~\cite{leon_bell-state_2021, gilbert_-demand_2023}. Consistent Rabi speeds are further favorable for global control protocols \cite{hansen_pulse_2021, seedhouse_quantum_2021}. Qubit driving frequencies are limited by the distance of the microwave antenna to the quantum dots compared to similar devices \cite{steinacker_bell_2025, huang_high-fidelity_2024}. Further, the power on the antenna was kept low due to the observation of Larmor frequency shifts as a function of the driving amplitude likely stemming from heating effects \cite{undseth_hotter_2023-1, noauthor_optimized_nodate, tanttu_assessment_2024}.
\change{
    The reduced Rabi speeds compared to similar devices \cite{huang_high-fidelity_2024, steinacker_industry-compatible_2025, jones_mid-circuit_2026} are likely attributable to the antenna design and distance from the qubit array, rather than to an inherent limitation of the qubits themselves.
}

\change{
    Temporal ensemble coherence times $T_2^*$, summarized in Fig.~2c are measured within a $\sim$\SI{4}{\min} time window by performing a Ramsey-type experiment followed by a state projection along all directions on the Bloch-sphere to calculate the state purity.
    The qubits provide up to \SI{41(2)}{\micro \second} of coherence, with an average of \SI{21(9)}{\micro\second}.
    Hahn-echo coherence times $T_2^{\mathrm{Hahn}}$ reach up to \SI{1.31(4)}{\milli \second}, with an average of \SI{0.7(4)}{\milli \second}, as shown in Fig.~2d.
    These values and variations are comparable to those reported for devices fabricated using a similar foundry process \cite{steinacker_industry-compatible_2025}, and exceed results from devices fabricated in academic facilities \cite{huang_high-fidelity_2024, jones_mid-circuit_2026}.
}

Reliably controlling electron exchange within a given DQD is not trivial with the given top-gate layout due to dots forming underneath the barrier gate [see Fig.~\ref{fig:mega_figure}c)iv and Supplementary Figure 5a] or by significant lateral shifts of the electrons spatial distribution [see right half of Fig.~\ref{fig:mega_figure}c)i] with increasing voltage on the barrier gates.
By loading a sufficient number of electrons into the  P1--P2 sub-system, we found an occupation providing a continuous range of J1 voltages without causing charge movements.   
\change{
    The remaining DQDs did not show exchange branching of their qubits electron spin resonance spectrum before one of the electrons tunneled to another spurious quantum dot.
    }
To characterize the CZ gate, we prepare one qubit in a superposition state $\ket{\psi} = \frac{1}{\sqrt{2}}(\ket{\uparrow_1} + \ket{\downarrow_1})\otimes\ket{\downarrow_2}$ and ramp to a certain gate voltage on J1 as well as detuning $\epsilon_{\text{P1-P2}}$ for a fixed wait time of \SI{1}{\micro \second}.
The qubit state is projected along the positive and negative x- and y-axis of the Bloch-sphere, using $\pm X_{\pi/2}$ and $\pm Y_{\pi/2}$ single qubit to determine the qubit's phase accumulation. This so called 'finger-print' measurement is shown in Fig.~\ref{fig:exchange}a.

Despite the non-uniform geometry of the quantum dots charge stability, the phase coherence is well preserved, hinting at low charge noise in the vicinity of the qubits~\cite{huang_fidelity_2019}. The finger-print map further reflects the bent structure of inter-dot electron crossings and even suggests a diagonal intra-dot transition above \SI{150}{\milli \volt} resulting in a sudden speed up of exchange oscillations \cite{steinacker_bell_2025}.
This measurement is repeated without applying a gate detuning $\epsilon_{\text{P1-P2}}$ while varying the wait time at the exchange voltage as shown in Fig. \ref{fig:exchange}b. The inset displays the exchange speed for each barrier gate voltage and an exponential turn on of 33.69(1) dec/V in qubit exchange is determined. 

Fig. \ref{fig:exchange}c shows the tuning of the phase adjustment of the CZ gate over up to 38 gate repetitions. Black and orange horizontal line cuts are shown in Fig. \ref{fig:exchange}d.
\change{
    The barrier gate voltage is increased via a square pulse for each CZ gate repetition, further pulse engineering can still improve the repeatability of the phase rotation.
    }

\change{
    Qubit exchange in the remaining DQDs did not turn on with barrier gate voltage, except for DQD P7--P8, which showed a sudden increase in exchange interaction as shown in Supplementary Figure 5b.
    }
    It is possible that with even higher electron occupancies the valence electrons wave functions overlap is increased to facilitate smooth qubit exchange \cite{leon_bell-state_2021}.
\change{
    Devices from a similar foundry process with reduced gate size and pitch, as reported in Ref.~\cite{steinacker_industry-compatible_2025}, have shown reliable and high-fidelity exchange control across all devices.
    Similarly, Ref.~\cite{jones_mid-circuit_2026} demonstrates that smaller gate pitches in a university-fabricated device enable reliable exchange among all neighboring qubits.
}
Albeit this is not a demonstration of the scalability of two qubit gate tuning, it shows low electrical noise being present in devices from this fabrication process \cite{reed_reduced_2016, steinacker_industry-compatible_2025}.

\ifpreprint
\begin{figure}
    \includegraphics[width=1.0\linewidth, angle = 0]{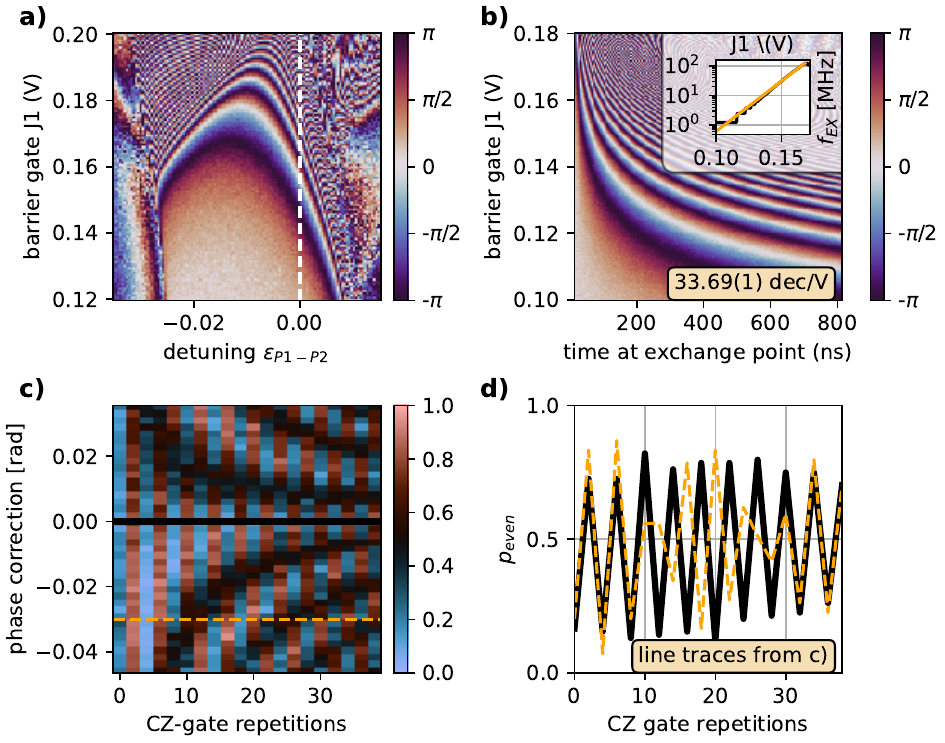}
    \caption{
        \textbf{Two qubit exchange in P1-P2.}
        \textbf{a)} Exchange decoupled `fingerprint' map at a fixed exchange period of \SI{1}{\micro \second} for barrier gate voltage J1 vs P1-P2 detuning voltage. 
        \textbf{b)} Exchange oscillations vs barrier gate voltage J1. The scan is performed along the white dotted line in a).
        \textbf{c)} Phase calibration of controlled-phase gate (CZ).
        \textbf{d)} Line cuts along black and orange line in c).
    }
    \label{fig:exchange}
\end{figure}
\fi

\section{Discussion}
\noindent



This work presents the coherent operation of eight silicon spin qubits in a \SI{300}{\milli \meter} foundry-fabricated linear quantum dot array. All eight qubits were successfully tuned and individually addressed, exhibiting coherence times, consistent with or close to state-of-the-art \cite{veldhorst_addressable_2014, steinacker_industry-compatible_2025}.
Double quantum dot cells, as they are calibrated in this work, break down the tuning complexity into $N$/2 units. To verify this approach, entangling gates among qubit pairs remain to be demonstrated. This was not easily feasible here, given the gate dimensions of the device and the expected shape of the electrons' wave functions \cite{cifuentes_bounds_2024}. Increasing the number of electrons captured in a dot increases the size of the qubits wavefunction \cite{leon_coherent_2020, leon_bell-state_2021} and thus the qubits potential to exhibit Heisenberg exchange with its neighbors. However, the higher charge occupancy leads to a tradeoff between expanded tuning complexity, lower charging energies and low excited state energies \cite{leon_coherent_2020, leon_bell-state_2021}.

One dimensional qubit arrays are a starting point to develop synchronous control techniques, giving insight into qubit statistics, and potentially enabling initial error correction codes \cite{jones_logical_2018}. Scalable, fault-tolerant spin qubit based error mitigation routines require higher-order qubit connectivity like bi-linear \cite{ha_two-dimensional_2025, siegel_towards_2024} or sparse 2D arrays \cite{boter_spiderweb_2022}.
\change{
    A dielectric resonator should be used for qubit driving in the higher dimensional architectures, \cite{vahapoglu_coherent_2022, vahapoglu_single-electron_2021} and global driving scheme \cite{hansen_implementation_2022, hansen_entangling_2024, seedhouse_quantum_2021} to scale single qubit driving.
    }
Demonstrating on-demand control over more complex multi quantum dot structures will enable the investigation of many proposed error correction schemes that are the foundation of a large scale quantum computer.
\change{
    Looking forward, continued progress in fabrication, including refined lithographic techniques, will be necessary to reduce feature sizes while preserving yield and uniformity, enabling robust qubit confinement and reliable exchange control in larger-scale devices.
}


\section{Methods}

\subsection{Fabrication}
\noindent
The device was fabricated at imec using a  300-mm wafer process engineered for low charge noise, high uniformity, and qubit-specific integration with a 90 nm gate pitch\cite{elsayed_low_2024, li_flexible_2021}. Optical and electron-beam lithography were combined to balance throughput with patterning precision. The starting material was an epitaxially grown silicon layer isotopically purified to a residual $^{29}$Si concentration of 400 ppm. Thermal oxidation of the silicon surface yielded the Si/SiO$_2$ interface where the electron spin qubit is hosted, with the interface quality maintained through dedicated cleaning steps. A triple-layer overlapping polysilicon gate stack was defined by electron-beam lithography and dry etching, with adjacent gates isolated by thin interstitial high-temperature oxide. Highly doped polysilicon was used to minimize interface strain at cryogenic temperatures. An aluminum strip line antenna above the qubit array was integrated to enable electron spin resonance control of individual spin states.

\subsection{Measurement}
\noindent
The device was cooled in an Oxford Instruments Kelvinox 400HA dilution refrigerator. An Oxford Instruments Mercury IPS controled the internal vector magnet providing a static magnetic field of \SI{0.5}{\tesla}, chosen to place the qubit Larmor frequencies in a regime where the on-chip stripline antenna operates efficiently.
D.C. bias voltages were applied through filtered lines (bandwidth $\leq$ \SI{30}{\hertz}) via a Q-Devil QDAC-II. A Quantum Machines OPX1000 served as the central controller for pulse generation, with a  \SI{1}{\nano\second} sampling time on LF-FEM baseband modules. Voltage pulses were combined with D.C. offsets at room temperature using custom linear bias combiners, and delivered through lines with a \SI{50}{\mega\hertz} bandwidth, imposing a minimum rise time of \SI{20}{\nano\second} that was mitigated by smoothing fast pulse edges. Microwave excitation up to \SI{10.5}{\giga\hertz} was provided by the MW-FEM module integrated within the OPX1000, which also synthesized the in-phase, quadrature, and pulse modulation waveforms. The dedicated Keysight E8267C vector signal generator was used to synthesize pulses up to \SI{20}{\giga\hertz} with baseband in-phase and quadrature pulses from an OPX LF-FEM module.
Charge readout was performed via two SETs in d.c. mode, placed on the lateral ends of the qubit array. The sensor current was integrated over $t_{int} =$ \SI{100}{\micro\second}, converted to a voltage with a room-temperature transimpedance amplifiers (Basel SP983c), and digitized by the OPX1000.

\section{Data Availability}
\noindent
The data supporting this work is available in a Zenodo repository (\href{https://zenodo.org/records/19583373}{19583373}) as a QCoDeS database.

\section{Code Availability}
\noindent
All results were acquired using a custom open-source measurement framework, with data analysis performed using the associated inspection tools \href{https://github.com/andncl/arbok_driver}{arbok-driver} and \href{https://github.com/andncl/arbok_inspector}{arbok-inspector}.

\bibliography{8Qbib}

\section*{Acknowledgments}
\noindent 
We thank A. Torgovkin for assistance with the cryogenic set-ups, S. Bartee and A. Klopper for feedback on the manuscript, and A. Dickie for help with cryogenic cables.

\section{Funding Statement}
\noindent
We acknowledge support from
the Australian Research Council (Grant No. FL190100167) and the U.S. Army Research Office
(W911NF-23-10092). P.S. and A.N. acknowledge support from the Sydney Quantum Academy
N.D.S. and K.W.C. are
recipients of an Australian Research Council Industrial Fellowship (Project Nos. IE240100252
and IM230100396) funded by the Australian Government.

\section{Author Contributions}
\noindent 
A.N. conducted the experiments under T.T. and A.S.D.’s supervision and with input from N.D.S., A.S., A.L., C.H.Y., P.S., J.D.C., M.K.F., S.S., E.V. and C.C.E.
The imec team, consisting of S.K., J.J., Y.C., S. Baudot, Y.S., R.L., C.G., B.R., S. Beyne and D.W. and led by K.D.G., developed the 300 mm spin-qubit process, fabricated the device and performed an initial electrical device screening at wafer-scale. W.H.L., K.W.C. and F.E.H. packaged the device for measurements.
A.N. wrote the paper with the input of all authors.

\section{Competing Interests}
\noindent 
A.S.D. is chief executive officer and a director of Diraq Pty Ltd. N.D.S,
W.H.L., T.T., M.K.F., S.S., J.D.C., E.V., F.E.H., K.W.C., A.L., C.H.Y., A.S., C.C.E. and A.S.D. declare
equity interest in Diraq. The other authors declare no competing interests.

\ifpreprint 

\FloatBarrier
\clearpage
\onecolumngrid
\appendix
\setcounter{page}{1}
\pagenumbering{roman}
\begin{center}
{\LARGE \textbf{Supplementary Information}}
\end{center}

\begin{center}
{Eight-Qubit Operation of a 300 mm SiMOS Foundry-Fabricated Device}
\end{center}
\FloatBarrier

\begin{suppfigure*}
    \includegraphics[width=0.8\textwidth, angle = 0]{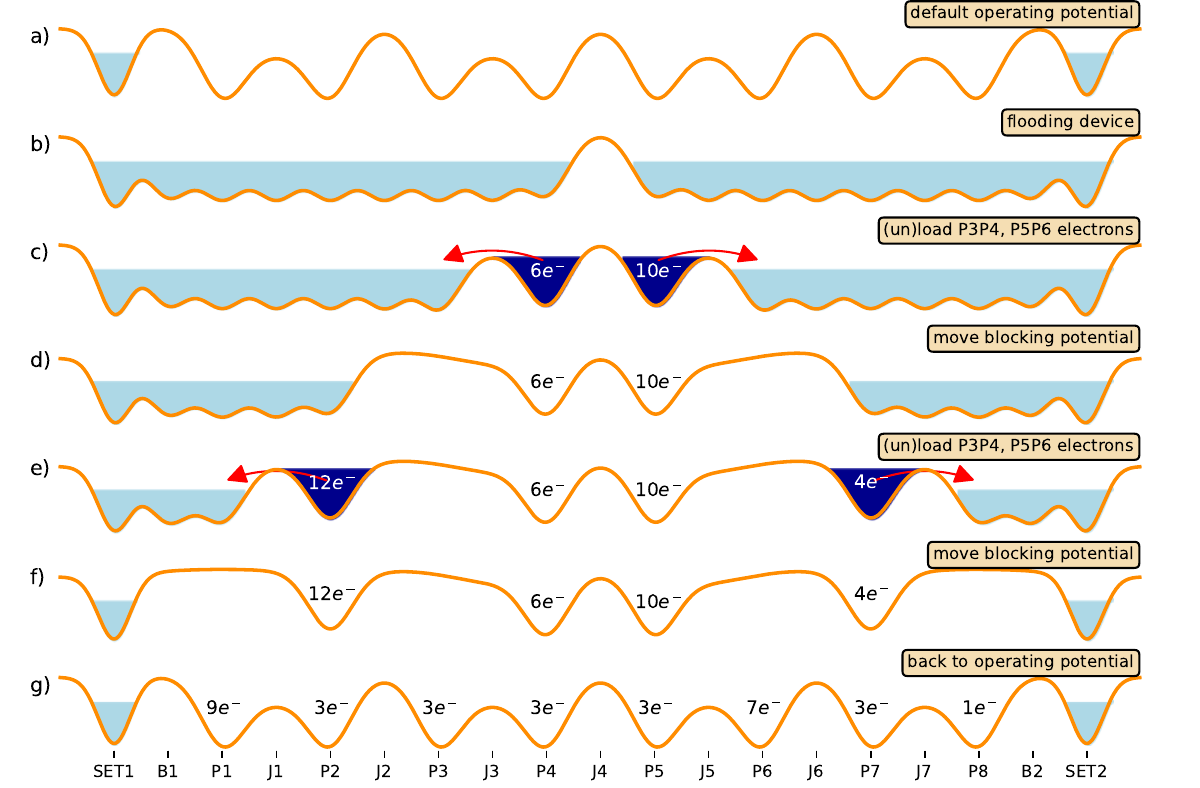} 
    \caption{
        \textbf{Electron loading sequence of the entire device.} 
        \textbf{a)} Initial DC operating potential.
        \textbf{b)} Flooding the device from both sides with a 2-dimensional electron gas from the SET islands symmetric around J4.
        \textbf{c)} Set calibrated loading voltages (P4, P5) and barrier voltages (J4, J5) to reduce the Fermi sea to the desired integer of electrons.
        \textbf{d)} Electrons are trapped to central dots by applying blocking potentials (J2, P3, J3 and J5, P6, J6).
        \textbf{e)} Similar to (c) but in lateral dots, loading voltages (P2 and P7) and barrier voltages (J1 and J7).
        \textbf{f)} Similar to (d) blocking potentials to trap electrons under P2 and P7 while pushing out the Fermi sea.
        \textbf{g)} Returning to the initial DC operating potential.
    }
    \label{supp:loading_sequence}
\end{suppfigure*}

\begin{suppfigure*}
    \includegraphics[width=0.9\textwidth, angle = 0]{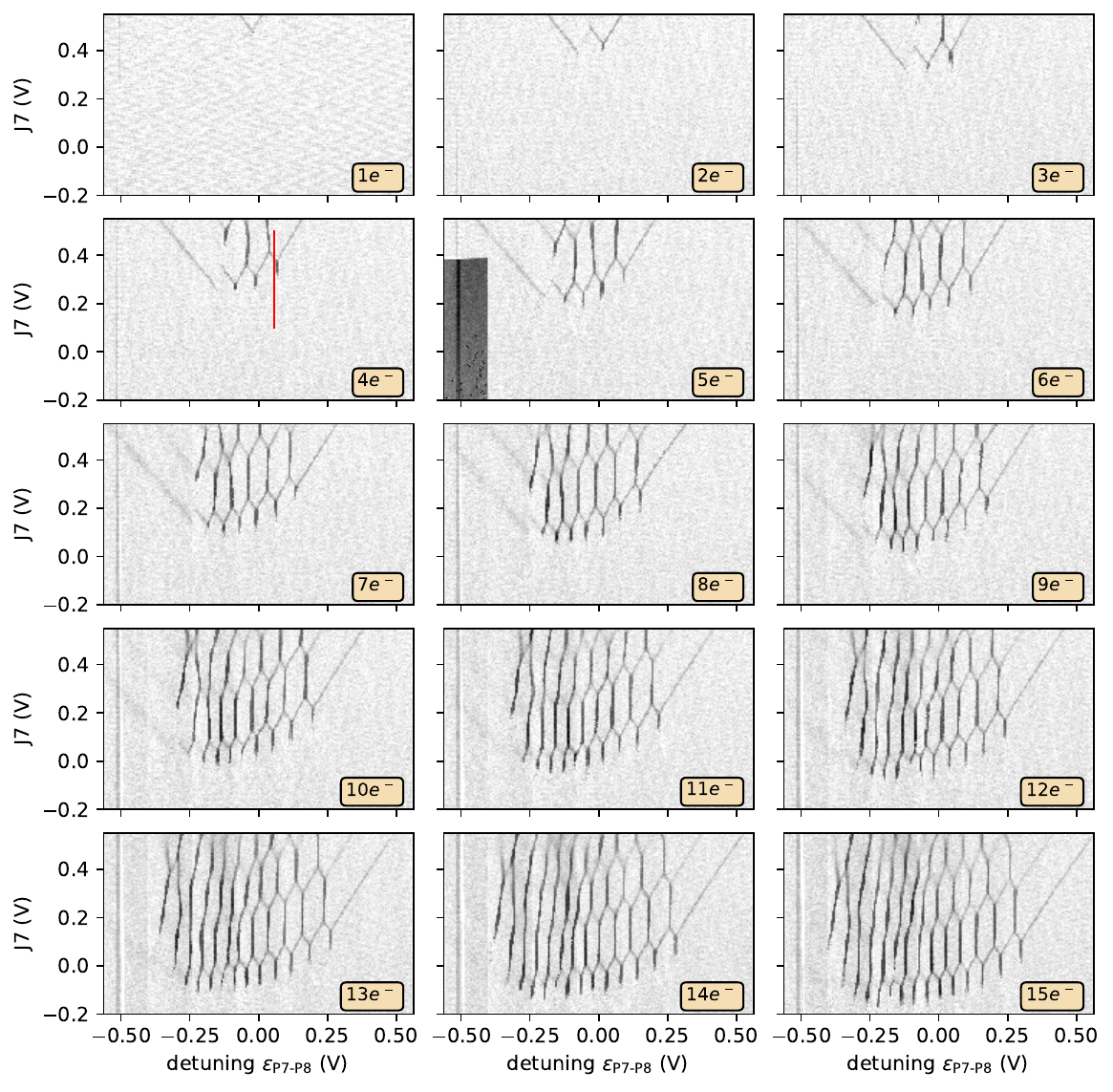} 
    \caption{
        \textbf{Isolated mode stability maps of P7P8 DQD | }
        Electrons are incrementally loaded into the double dot using the technique shown in Supplementary Figure \ref{supp:loading_sequence}. 
        The higher the voltage of P7 during the sequence stop shown in Supplementary Figure \ref{supp:loading_sequence}e, the more electrons will be captured in P7--P8. The loading voltage decreases with an increased number of electrons per DQD~\cite{leon_coherent_2020}. Inter-dot transitions (vertical) are extending, following charge occupancy numbers, thus tunnel coupling among DQDs before forming a quantum dot under the intermediate J-gate is enhanced as more electrons are accumulated.
        The red line in the four electron maps indicates the scan shown in Supplementary Figure \ref{supp:ex_turonon}a.
    }
    \label{supp:e_loading}
\end{suppfigure*}

\begin{supptable}[h!]
\centering
\begin{tabular}{ | c | c | c | c |}
\hline
\textbf{P1P2} & \textbf{P3P4} & \textbf{P5P6} & \textbf{P7P8}\\
\hline
\cellcolor{yellow} (3-3) & \cellcolor{green} (3-3) & \cellcolor{red!30} (3-3) & \cellcolor{green} (3-1) \\
\cellcolor{yellow} (5-3) & \cellcolor{yellow} (5-3) & \cellcolor{yellow} (3-5) & \cellcolor{yellow}(3-3)\\
\cellcolor{green}(9-3) &                            & \cellcolor{green} (3-7) & \cellcolor{red!30}(5-3)\\
\cellcolor{red!30} (13-3) &  & & \\
\hline
\end{tabular}
\caption{Charge configurations examined in this measurement campaign for all double quantum dots. Green highlights indicate the used charge configuration per double dot in the main figure 1. Yellow highlights mark charge configurations that allowed the tuning of qubits which showed Rabi oscillations with lower Q-factors. Fig. \ref{supp:other_charge_configs} shows some of the qubit results for the yellow charge configurations. Charge configurations in red yielded no qubits.}
\label{tab:other_change_configs}
\end{supptable}

\begin{suppfigure*}
    \includegraphics[width=\textwidth, angle = 0]{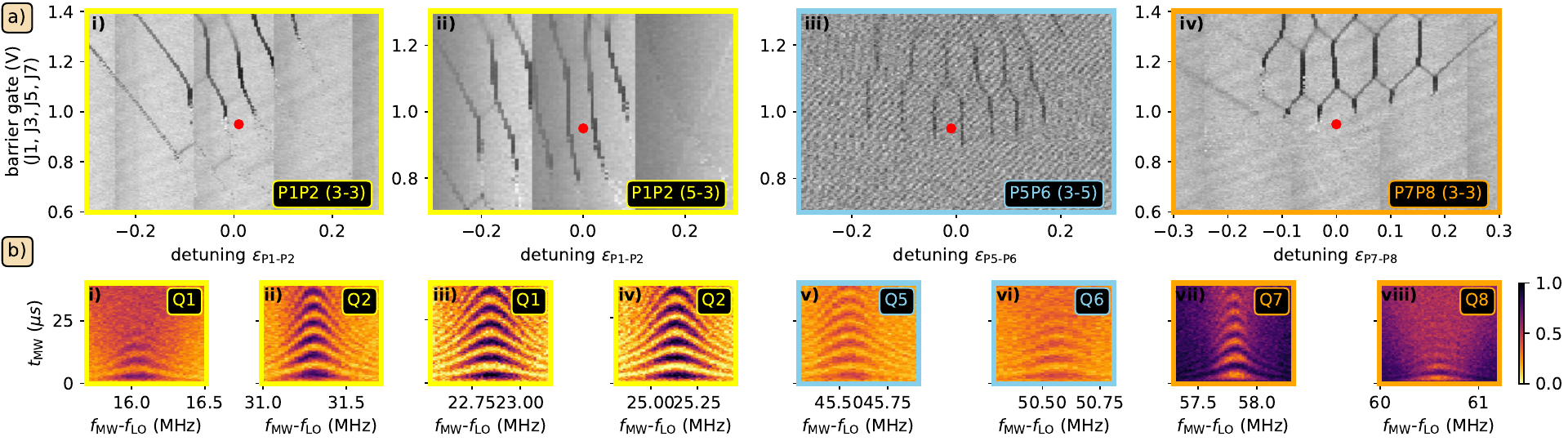} 
    \caption{
        \textbf{Charge stability maps and Rabi-chevron measurement for different electron numbers |} 
        Red circles mark the charge
configuration used for the measured qubits.
        \textbf{P1-P2 (3-3):} a)i, b) i,ii
        \textbf{P1-P2 (5-3):} a)ii, b) iii,iv
        \textbf{P5-P6 (3-5):} a)iii, b) v,vi; Rabi chevrons were measured directly via SET2 without electron cascade, hence the low visibility
        \textbf{P7-P8 (3-3):} a)iv, b) vii,viii
    }
    \label{supp:other_charge_configs}
\end{suppfigure*}

\begin{suppfigure*}
    \includegraphics[width=0.9\textwidth, angle = 0]{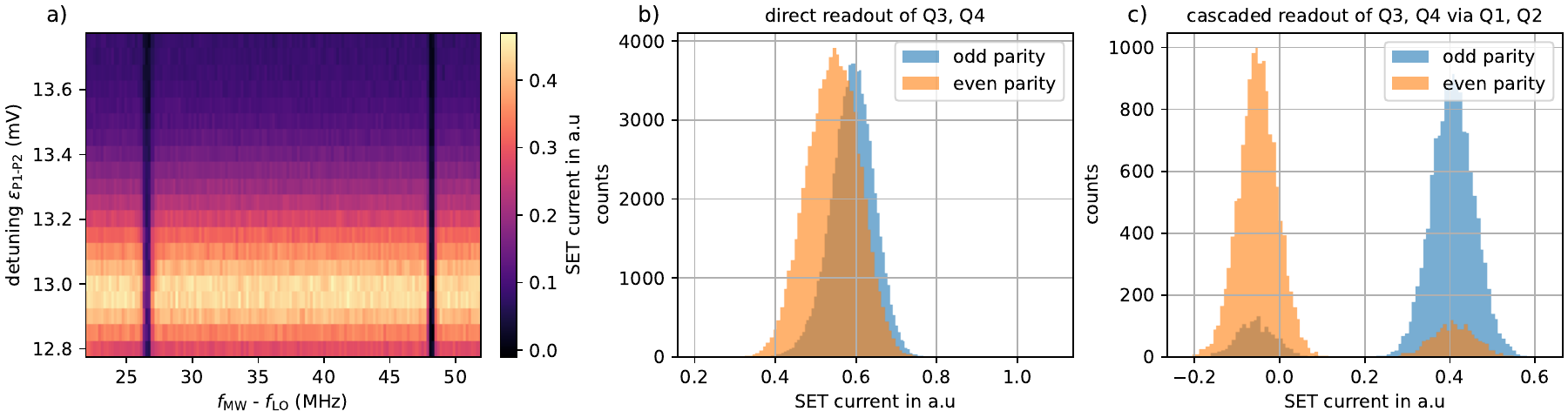} 
    \caption{
        \textbf{Calibration of cascaded readout | } 
        \textbf{a)} Difference in visibility of an ESR measurement when changing the detuning $\epsilon_{P1-P2}$. The high contrast region corresponds to a cascaded electron transition from P1 to P2
        \textbf{b)} Histograms of an ESR measurement using direct readout of P3P4 double dot
        \textbf{c)} Histograms of an ESR measurement using cascaded readout of P3P4 double dot via P1P2 cascading
    }
    \label{supp:cascading_calibration}
\end{suppfigure*}

\begin{suppfigure*}
    \includegraphics[width=0.65\textwidth, angle = 0]{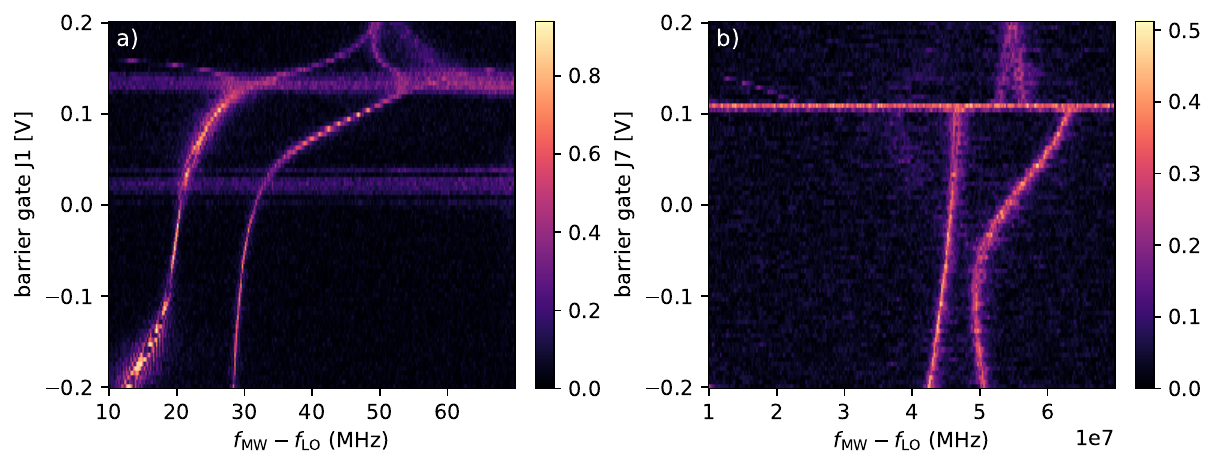} 
    \caption{
        \textbf{Two qubit exchange | } ESR measurement vs barrier gate voltage in DQDs. Horizontal splitting in qubits ESR frequencies at high barrier gate voltage corresponds to their exchange coupling.
        \textbf{a)} P1P2, continuous exponential turn on of exchange through Heisenberg interaction at (9,3) charge configuration
        \textbf{b)} P7P8, sudden turn of exchange via shuttling to intermediate J-dot under J7 at (1,3) charge configuration. The corresponding charge stability map with 4 electrons can be seen in Supplementary Figure \ref{supp:e_loading}. The scan axis of the measurement shown here is indicated by the red line
    }
    \label{supp:ex_turonon}
\end{suppfigure*}

\begin{suppfigure*}
    \includegraphics[width=0.65\textwidth, angle = 0]{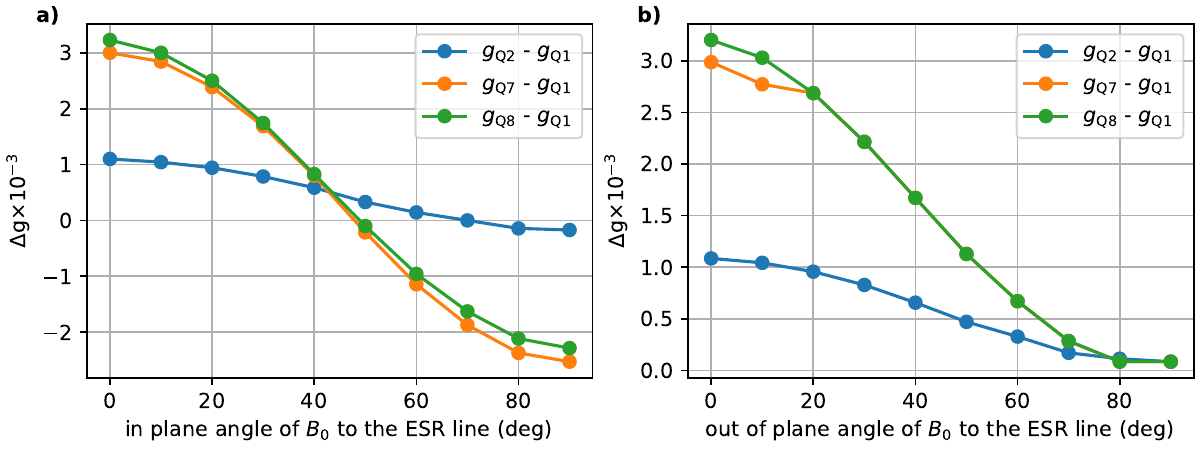}
    \caption{
        \textbf{Qubit Larmor frequencies vs magnetic field angle $B_0$| }
        \textbf{a)} rotation $\psi$ in the plane spanned by the qubit array and the normal to the schematic shown in Fig 1a. Rotation from [110] to [1$\bar{1}$0] ($45^{\circ}$)
        \textbf{b)} rotation $\theta$ in the plane depicted in schematic \ref{fig:mega_figure}a. Both angles being 0 results in $B_0$ pointing parallel to the qubit array. Rotation from [110] to [001]
    }
    \label{supp:b_field_rot}
\end{suppfigure*}

\begin{suppfigure*}
    \includegraphics[width=0.5\textwidth, angle = 0]{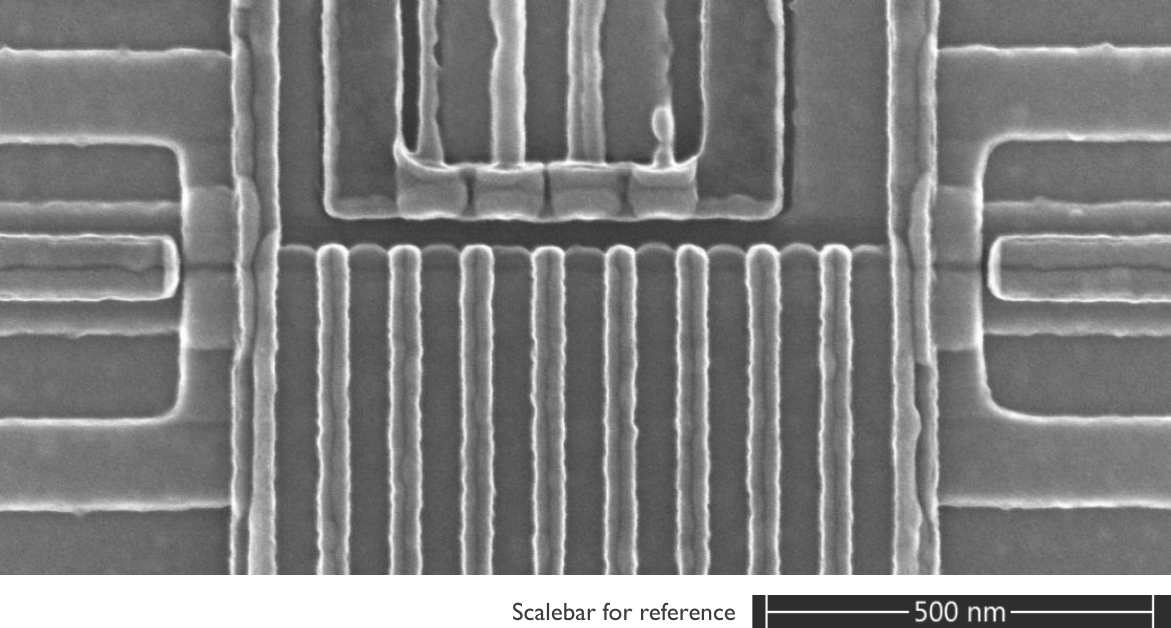}
    \caption{
        \textbf{SEM image of a similar device} \cite{li_flexible_2021}
    }
    \label{supp:SEM}
\end{suppfigure*}

\twocolumngrid

\fi

\end{document}